\documentclass{kapproc} 
\let\footnote\savefootnote
\let\footnotetext\savefootnotetext 
\let\footnoterule\savefootnoterule 
\kluwerbib

\RequirePackage{graphicx}%
\RequirePackage{epsf}%
\usepackage{amsmath}

\begin{document}

\articletitle{Hydrodynamical Simulations of the IGM at High Mach Numbers}

\author{Trac, H.$^{1,2}$, Zhang, P.$^{1,2}$, \& Pen, U.$^2$}

\affil{$^1$ Department of Astronomy and Astrophysics, University of Toronto, ON M5S 1A7, Canada\\ $^2$ Canadian Institute for Theoretical Astrophysics, 60 St. George Street, Toronto, ON M5S 3H8, Canada}
\email{trac@cita.utoronto.ca}


\begin{abstract}
We present a new approach to doing Eulerian computational fluid dynamics that is designed to work at high Mach numbers encountered in hydrodynamical simulations of the IGM.  In conventional Eulerian CFD, the thermal energy is poorly tracked in supersonic bulk flows where local fluid variables cannot be accurately separated from the much larger bulk flow components.  We describe a method in which local quantities can be directly tracked and the Eulerian fluid equations solved in a local frame moving with the flow.  The new algorithm has been used to run large hydrodynamical simulations on a $1024^3$ grid to study the kinetic SZ effect.  The KSZ power spectrum is broadly peaked at $l\sim10^4$ with temperature fluctuations $\sim1\mu$K.
\end{abstract}

\newcommand{\bsk}{\boldsymbol{k}}
\newcommand{\bsp}{\boldsymbol{p}}
\newcommand{\bsv}{\boldsymbol{v}}
\newcommand{\bsx}{\boldsymbol{x}}


\section{Introduction}

Reionization of the universe has left an abundance of free electrons which can be used as a direct probe of the intergalactic medium (IGM).  Cosmic microwave background (CMB) photons are scattered by free electrons with thermal and bulk motions via inverse Compton or Doppler scattering.  These two processes produce secondary temperature fluctuations in the CMB and are known as the thermal and kinetic Sunyaev-Zeldovich (SZ) effects, respectively (Sunyaev \& Zeldovich 1972, 1980).  The SZ effects can be used as a robust probe of the IGM since they provide a complete and unbiased inventory of all free electrons.

The SZ effects are the dominant secondary CMB anisotropies on arcminute scales.  Recent hydrodynamical simulations of the thermal SZ (TSZ) effect predict temperature fluctuation in the tens of $\mu$K (\cite{sbp00,zpw02}), while the kinetic SZ (KSZ) effect produces $\mu$K fluctuations (\cite{swh01,zpt02}).

Upcoming CMB experiments such as AMiBA, SZA, and Planck have the potential to measure both these effects.  The TSZ and KSZ have different frequency dependences, allowing them to be separated observationally.  At a frequency $\nu\approx210$ GHz, the TSZ effect drops out and the KSZ effect becomes the dominant secondary anisotropy.  While the KSZ distortion is smaller in amplitude, it is a very clean problem to solve since the effect has a simple dependence on the ionized gas momentum.

To compliment the observational data, theoretical modeling is needed to obtain a more quantitative understanding of the KSZ effect and its potential as a probe of the IGM.  There is still much discrepancy amongst theoretical predictions based on analytical and numerical methods.  The focus of this paper will be on the development of theoretical tools which will enhance our understanding of the IGM.

\section{Kinetic SZ Effect}

The KSZ effect results from the Doppler scattering of CMB photons off of free electrons with bulk motion.  The temperature fluctuation is given by
\begin{equation}
\frac{\Delta T}{T}=\int n_e\sigma_T\bsv\cdot\boldsymbol{\hat{n}}\ dl\ ,
\end{equation}
where $n_e$ is the electron density, $\bsv$ is the velocity field, and $\boldsymbol{\hat{n}}$ is a unit vector in the direction of the line of sight.  The normalized momentum field
\begin{equation}
\bsp\equiv(1+\delta)\bsv=\bsp_B+\bsp_E\ ,
\end{equation}
can be split into a rotational part $\bsp_B$ and an irrotational part $\bsp_E$.  It can be shown that only the rotational part $\bsp_B$ contributes to the KSZ effect.  Using Limber's approximation (\cite{lim54}), the CMB power spectrum amplitudes $C_l$ can be related to the power spectrum of momentum B mode field
\begin{equation}
P_B(k)\equiv\langle|\bsp_B(\bsk)|^2\rangle\ .
\end{equation}
The KSZ effect is a nonlocal problem and has contributions from both large-scale and small-scale power.  This makes it a challenging problem to numerically model because high resolution is needed for all scales.

\section{Eulerian Hydrodynamical Simulations}

At high redshifts, the evolution of the IGM is only a moderately nonlinear problem and it is tractable to model the physics with a high level of accuracy and confidence using hydrodynamical simulations.  Eulerian computational fluid dynamics (CFD) is ideal for simulating the KSZ effect because it provides high mass resolution which is needed everywhere.  Eulerian codes also have the advantage of being computationally very fast, memory efficient, and easy to implement in parallel.

One of the key challenges of the problem is simulating astrophysical fluid in the presence of high Mach numbers.  In cosmological simulations, there are large-scale velocity fields on the order of 1000 km/s and the typical sound speed in these bulk flows is $\sim$ 10 km/s.  At Mach numbers $\sim$ 100, the ratio of the thermal energy to the kinetic energy is $\sim 10^{-4}$.  In an Eulerian code, the standard practice is to calculate the thermal energy by subtracting the kinetic energy from the total energy, but this calculation is inaccurate at high Mach numbers.  The poor tracking of the thermal energy in supersonic bulk flows is known as the high Mach number problem in Eulerian CFD.  This is especially problematic when the temperature distribution is required to compute radiative cooling and heating rates.

\section{Moving/Advecting Cells Hydrodynamics}

We have developed an approach to doing Eulerian CFD that is designed to work at high Mach numbers. The velocity field of the fluid,
\begin{equation}
\bsv(\bsx)=\Delta\bsv(\bsx)+\bsv_g(\bsx)
\end{equation}
is decomposed into a local term $\Delta\bsv(\bsx)$ and a smoothed background term $\bsv_g(\bsx)$.  The local velocity is a peculiar velocity with respect to the Eulerian grid and the smoothed background velocity is associated with the velocity of the grid cells.  The total energy,
\begin{equation}
e(\bsx)=\Delta e(\bsx)+e_g(\bsx)\ ,
\end{equation}
can also be separated into a local term $\Delta e(\bsx)$ and a grid term $e_g(\bsx)$.  This decomposition allows us to keep track of the local quantities and remove the bulk flow component in the velocity field and the total energy.  The local total energy is defined by
\begin{equation}
\Delta e(\bsx)=\frac{1}{2}\rho\Delta v^2+\varepsilon\ ,
\end{equation}
where $\rho$ is the density field and $\varepsilon$ is the thermal energy.  Since the bulk flow component has been removed, the thermal energy and the local kinetic energy will be comparable.  The {\it MACH} code (\cite{tra02}) works by solving the fluid equations in a local frame where the fluid moves at $\Delta\bsv(\bsx)$.  Then the cells are advected at the grid velocity $\bsv_g(\bsx)$ using a Lagrangian-Eulerian remap technique.

\section{Hydrodynamical Simulations of the KSZ Effect}


\begin{figure}[th]
\epsscale{0.65}
\plotone{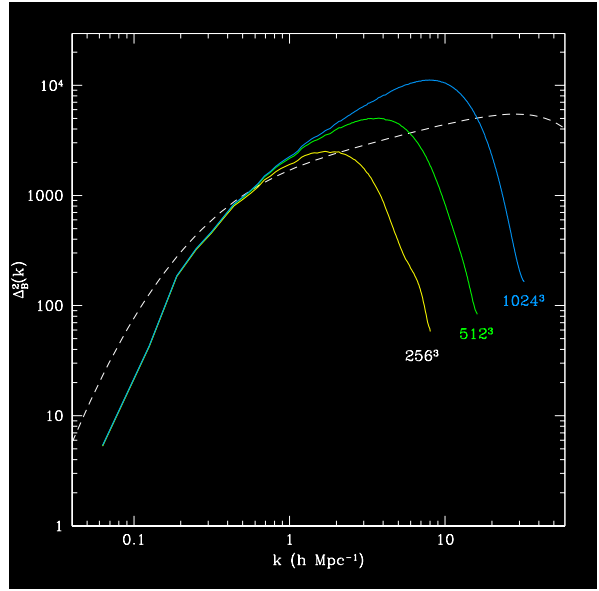}
\caption{The momentum B mode power spectra from simulations with different resolutions are given by the solid lines.  The dotted line is the Vishniac prediction from second-order perturbation theory.}
\label{fig:pb}
\end{figure}

\begin{figure}[th]
\epsscale{0.65}
\plotone{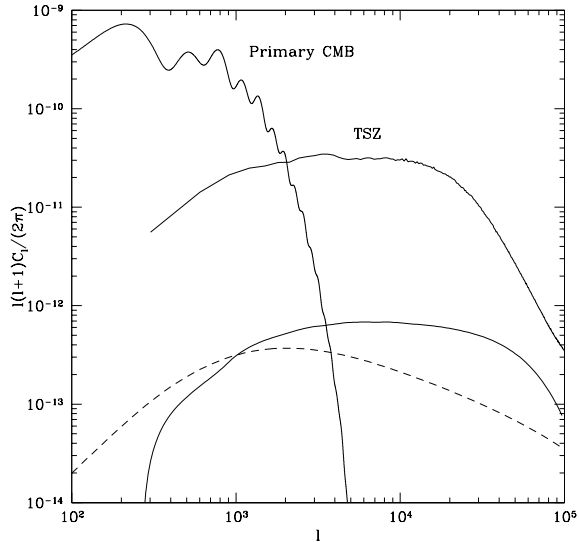}
\caption{The KSZ power spectrum from the $1024^3$ MACH simulation is compared with the TSZ power spectrum from a $512^3$ MMH simulation.  The dotted line is the Vishniac prediction for the KSZ effect.}
\label{fig:cl}
\end{figure}

We have run hydrodynamical simulations using the MACH code to study the KSZ effect in a $\Lambda$CDM universe.  We simulated a flat cosmological model with $\Omega_m=0.34$, $\Omega_\Lambda=0.66$, $\Omega_b=0.05$, $h=0.66$, and $\sigma_8=0.948$ in a 100 $h^{-1}$Mpc box.  Reionization is unformly and instantaneously imposed at a redshift of 10.

In Figure (\ref{fig:pb}) we plot the momentum B mode power spectra from simulations with different resolutions and compare them to the linear prediction from second-order perturbation theory (\cite{vis87}).  The simulations lose power on large scales due to the finite box size which prevents the capturing of large-scale velocity fields.  \cite{zpt02} show how the truncation of power due to a finite box size can be corrected.  On small scales, the numerical results have not converged but it is still evident that they show more power than the linear prediction.  Nonlinear analytical modeling of the Vishniac effect also shows an excess of power relative to the linear prediction (\cite{hu00,mf02}).  The truncation of power at the smallest scales is due to the finite resolution and gas pressure.

In Figure (\ref{fig:cl}) we plot the KSZ power spectrum from the $1024^3$ simulation and compare it with the TSZ power spectrum (\cite{zpw02}) from a $512^3$ simulation run with the Moving-Mesh Hydro (MMH) algorithm (\cite{pen98}).  The TSZ spectrum has a broad peak that spans $l\sim 5\times10^3 - 5\times10^4$ and has a maximum amplitude $\sim 10^{-12}$ which is $\sim30$ times smaller than the TSZ.  The maximum amplitude is consistent with the results from SPH simulations (\cite{swh01}), but the MACH results shows more power at large scales.  A code comparison program is planned to address the numerical discrepancies from different hydrodynamical algorithms and resolution.

\section{The Future of Cosmological Simulations}

We are currently working on projects to run very large cosmological simulations where both small-scale and large-scale effects can be probed simultaneously.  

One project involves running N-body simulations on distributed memory machines.  At the Canadian Institute for Theoretical Astrophysics, we have an 8-node cluster of Intel Itanium machines, where each node has 4 processors and 64 GB of memory.  With a total of 512 GB of distributed memory, we can run N-body simulations with 2000$^3$ particles on a 4000$^3$ grid.  

We are developing a two-level particle-mesh (PM) code to run on the Itanium cluster.  Two levels of grid will be used to calculate the gravitational force, which is split into a short-range and a long-range component.  The short-range force is calculated locally on a fine-grained mesh while the long-range force is done globally on a coarse-grained mesh.  The forces are computed using fast FFT's with the Intel IPP library.  A 512$^3$ FFT can be done in less than 10 seconds using one processor.  We estimate that one time step will take 10 minutes to run and a simulation with 500 time steps can be run in less than 4 days.

The second project is a unique niche in computing paradigms and it involves running out-of-core hydrodynamical simulations.  Out-of-core computation refers to the idea of using disk space as virtual memory and transferring data in and out of main memory at high I/O bandwidths.  We have a SCSI array with 96 disks which collectively provide 3.5 TB of virtual memory.  The disks are arranged in striped daisy chains to maximize the I/O performance.  With 3.5 TB of storage, we can run hydrodynamical simulations with 4000$^3$ cells.

\section{Summary}

A new hydrodynamical algorithm designed to solve the high Mach number problem encountered in Eulerian CFD has been applied to simulating the KSZ effect.  Detailed numerical results and discussions are presented in \cite{zpt02}.

Algorithms to run large N-body and hydrodynamical simulations on a $4000^3$ grid are currently in development.  An MPI version of the two-level PM N-body code will be made publicly available in the near future.



\end{document}